\documentclass[iop]{emulateapj}
\usepackage{graphicx,amsmath,amssymb}
\usepackage{natbib}
\usepackage{apjfonts}
\usepackage{multirow}

\begin{document}

   \title{Thermal emission in the early X-ray afterglows of GRBs: following the prompt phase to late times}

   \author{Mette~Friis}\affil{Centre for Astrophysics and Cosmology, Science Institute, University of Iceland, Dunhagi 5, 107 Reykjav\'ik, Iceland;\\\texttt{mef4@hi.is}}
\and
   \author{Darach~Watson}\affil{Dark Cosmology Centre, Niels Bohr Institute, University of Copenhagen, Juliane Maries Vej 30, DK-2100 Copenhagen \O, Denmark;\\\texttt{darach@dark-cosmology.dk}}
   
   \def\titlerunning{Thermal emission in X-ray afterglows of GRBs}

   \begin{abstract}

    Thermal radiation, peaking in soft X-rays, has now been
    detected in a handful of GRB afterglows and has to date been interpreted
    as shock break-out of the GRB's progenitor star.  We present a search
    for thermal emission in the early X-ray afterglows of a sample of
    \emph{Swift} bursts selected by their brightness in X-rays at early
    times.  We identify a clear thermal component in eight GRBs and track
    the evolution.  We show that at least some of the emission must come
    from highly relativistic material since two show an apparent
    super-luminal expansion of the thermal component.  Furthermore we
    determine very large luminosities and high temperatures for many of
    the components---too high to originate in a SN shock break-out.  Instead we
    suggest that the component may be modelled as late photospheric emission
    from the jet, linking it to the apparently thermal component observed
    in the prompt emission of some GRBs at gamma-ray and hard X-ray
    energies.  By comparing the parameters from the prompt emission and the
    early afterglow emission we find that the results are compatible with
    the interpretation that we are observing the prompt quasi-thermal
    emission component in soft X-rays at a later point in its evolution.

   \end{abstract}

   \maketitle

%
%
\section{Introduction\label{introduction}}

Gamma-ray bursts (GRBs) are extremely bright transient sources, completely dominating the gamma-ray sky for milliseconds to half an hour. The nature of these bursts was long unknown, but since discovering an association with type Ic supernovae (SNe) (e.g. \citealt{1998Natur.395..670G,2003Natur.423..847H,2003ApJ...591L..17S}), long GRBs (any burst lasting longer than two seconds) are known to be caused by the core-collapse of massive stars. While the afterglows have provided a lot of information over the past 15 years, there is still a lot of uncertainty associated with the prompt high energy radiation. Recent evidence has been found in some GRBs for an apparently thermal component in the early soft X-ray emission \citep{2006Natur.442.1008C,2011MNRAS.411.2792S,2011MNRAS.416.2078P}. This component is often, though not exclusively \citep[e.g.][]{2007MNRAS.382L..77G}, interpreted as originating from the SN shock-breakout, as associated SNe have been detected spectroscopically for GRB\,060218/SN2006aj \citep{2006Natur.442.1008C,2006Natur.442.1011P} and GRB\,100316D/SN2010bh \citep{2011MNRAS.411.2792S,2010GCN..10525...1W,2010arXiv1004.2262C,2011arXiv1111.4527B}, and photometrically for GRB\,090618 \citep{2011MNRAS.416.2078P,2011MNRAS.413..669C}. 

Interestingly, a weak second component was found to be required statistically in the combined late prompt/early afterglow spectra of a few very soft GRBs \citep{2008A&A...478..409M}, suggesting either the emergence of the afterglow or a thermal component. The clear evidence for individual X-ray thermal components has so far only been detected in low redshift bursts with SNe. \cite{2012MNRAS.427.2965S} searched for this thermal emission in the total sample of \emph{Swift} bursts, but removed all bursts with high redshift, as they argue that the thermal emission detected in these are at high risk of being false positives. We have analysed a sample of bright early X-ray afterglows for the presence of thermal-type emission as well, but without the redshift filter, as we see no reason to rule these out beforehand. We find several new apparent thermal components, including bursts with redshifts $z>1$. In section~\ref{observations} we present the sample and the analysis of the Burst Alert Telescope (BAT) and the X-Ray Telescope (XRT) data. In section~\ref{results} the results of our time-resolved spectroscopy is presented for the best fit time series of each burst. In section~\ref{discussion} we examine the implications of the physical parameters deduced from modeling the emission component.

A flat universe cosmology is assumed with $H_0=70$\,km\,s$^{-1}$\,Mpc$^{-1}$ and $\Omega_{\rm M} = 0.27$. 1\,$\sigma$ errors for each parameter have been used. When only upper limits are given, these are 3\,$\sigma$.
%
%
\section{Observational data and methods}\label{observations} 

The sample presented in this paper consists of the brightest bursts (as of
2011 December 20) in the \emph{Swift} online catalogue
\citep{2007A&A...469..379E}.  We have selected bursts with at least 20\,000
counts in the XRT Window Timing (WT) data as well as reliable spectroscopic
redshifts, to ensure good fit statistics and to be able to study the
evolution in time and the rest-frame properties of the afterglows.  Twenty
nine bursts fit these criteria, but the dataset for
GRB\,100906A had to be discarded as repeated extractions of the spectra
failed to produce reliable results, so the final sample size is twenty eight bursts (see
Table~\ref{tab:best}).

\begin{figure*}
\includegraphics[bb=18 210 575 590, width=0.475\textwidth]{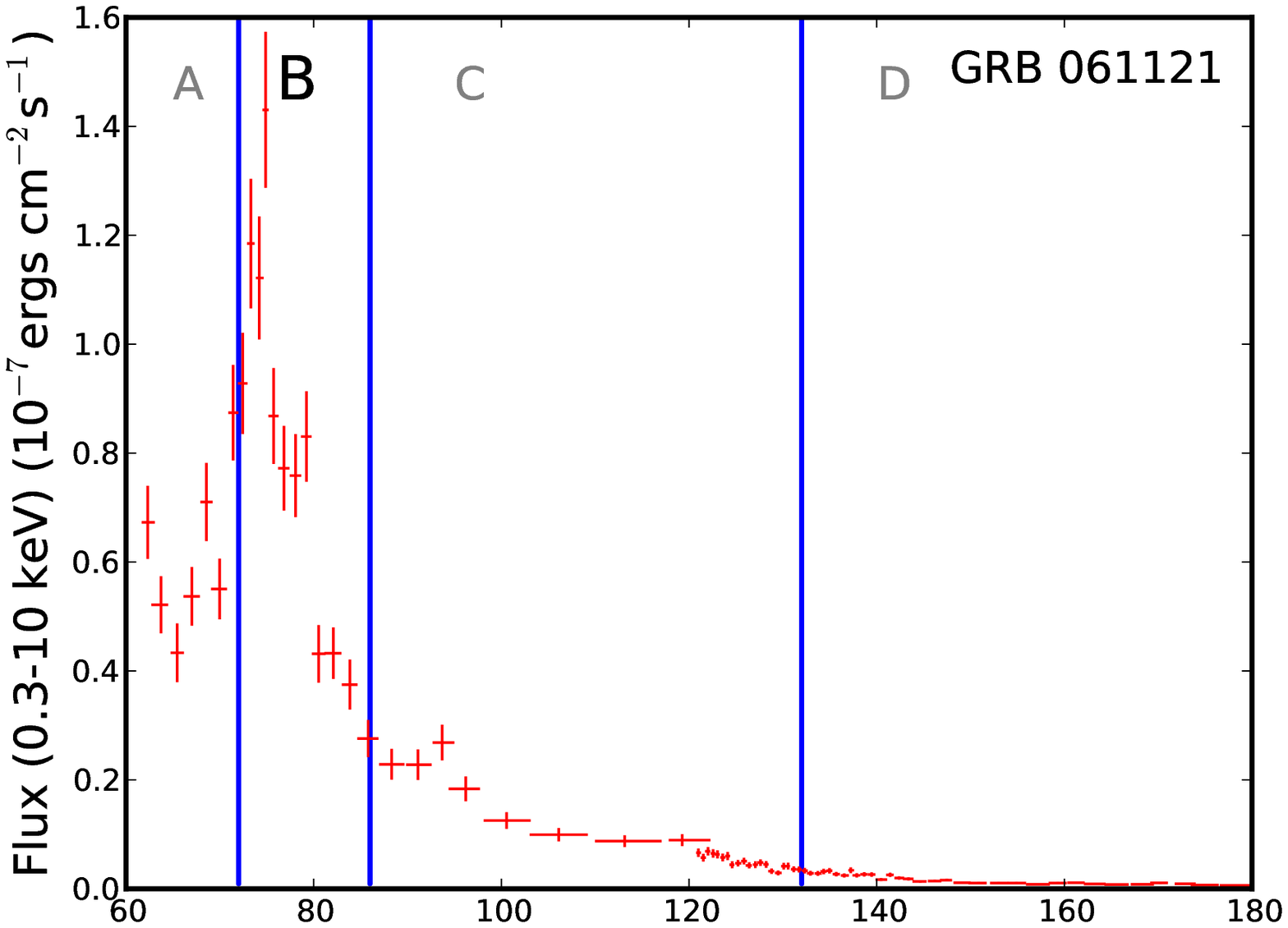}
\includegraphics[bb=18 210 575 590, width=0.475\textwidth]{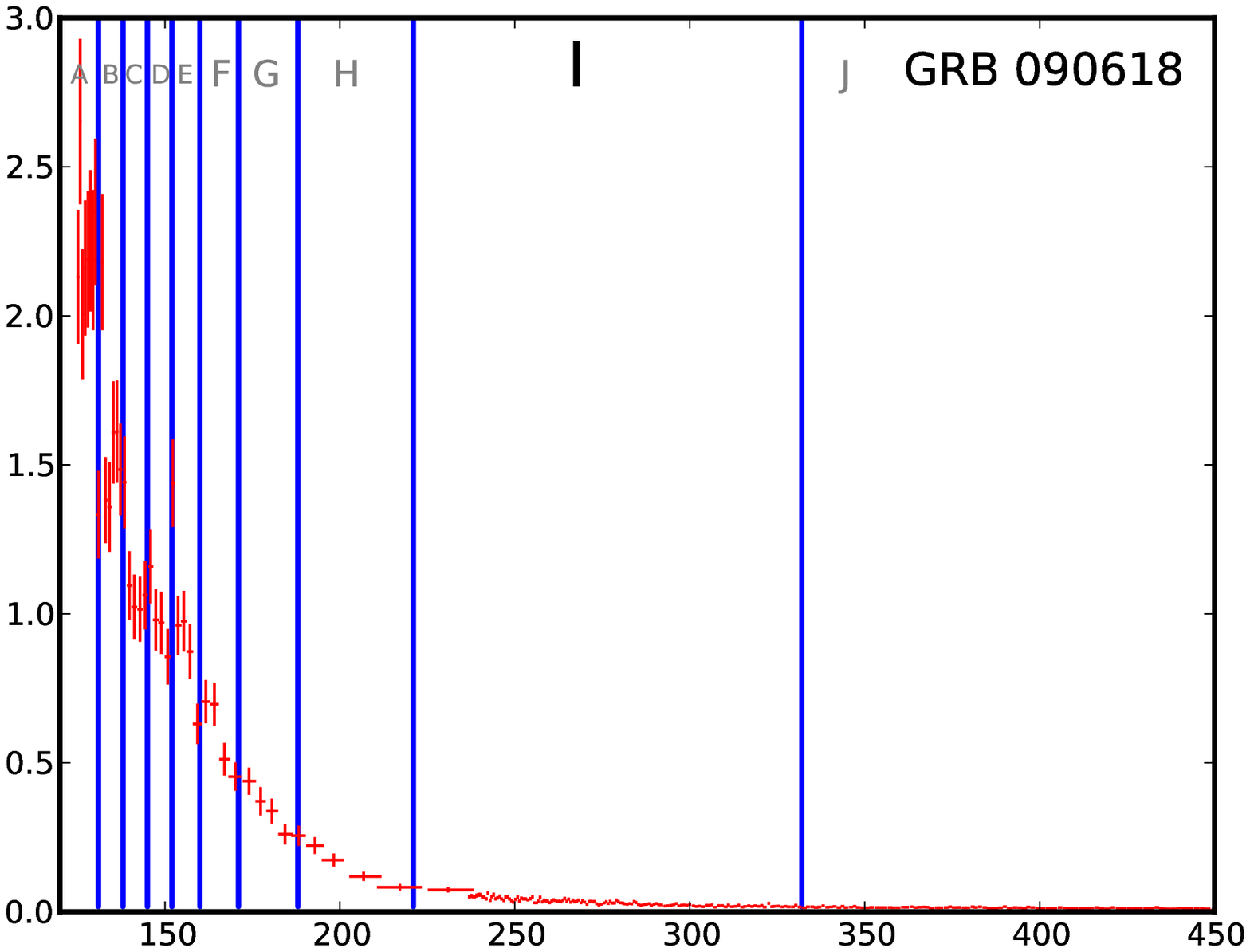}
\includegraphics[bb=18 210 575 609, width=0.475\textwidth]{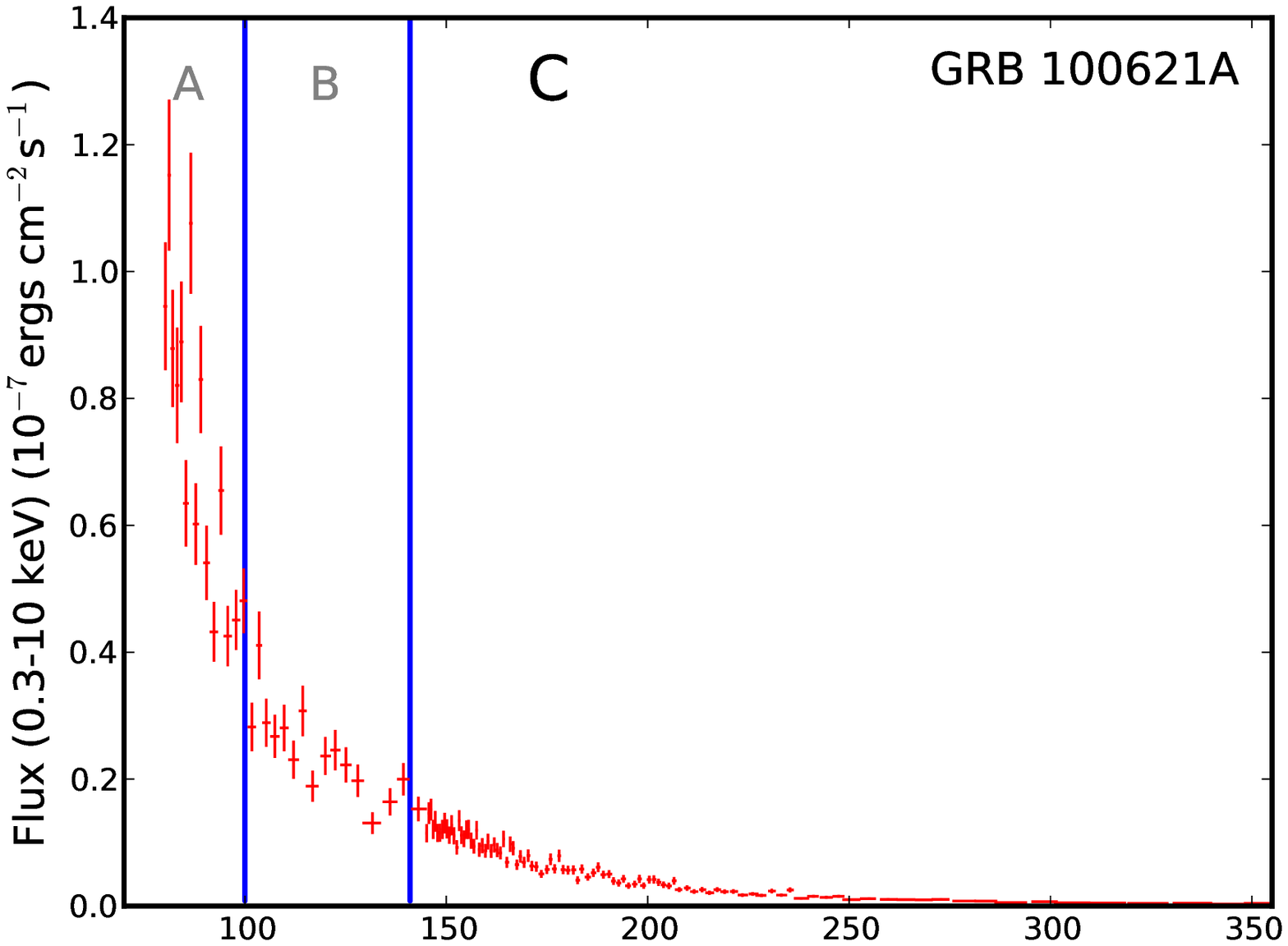}
\includegraphics[bb=18 210 575 609, width=0.475\textwidth]{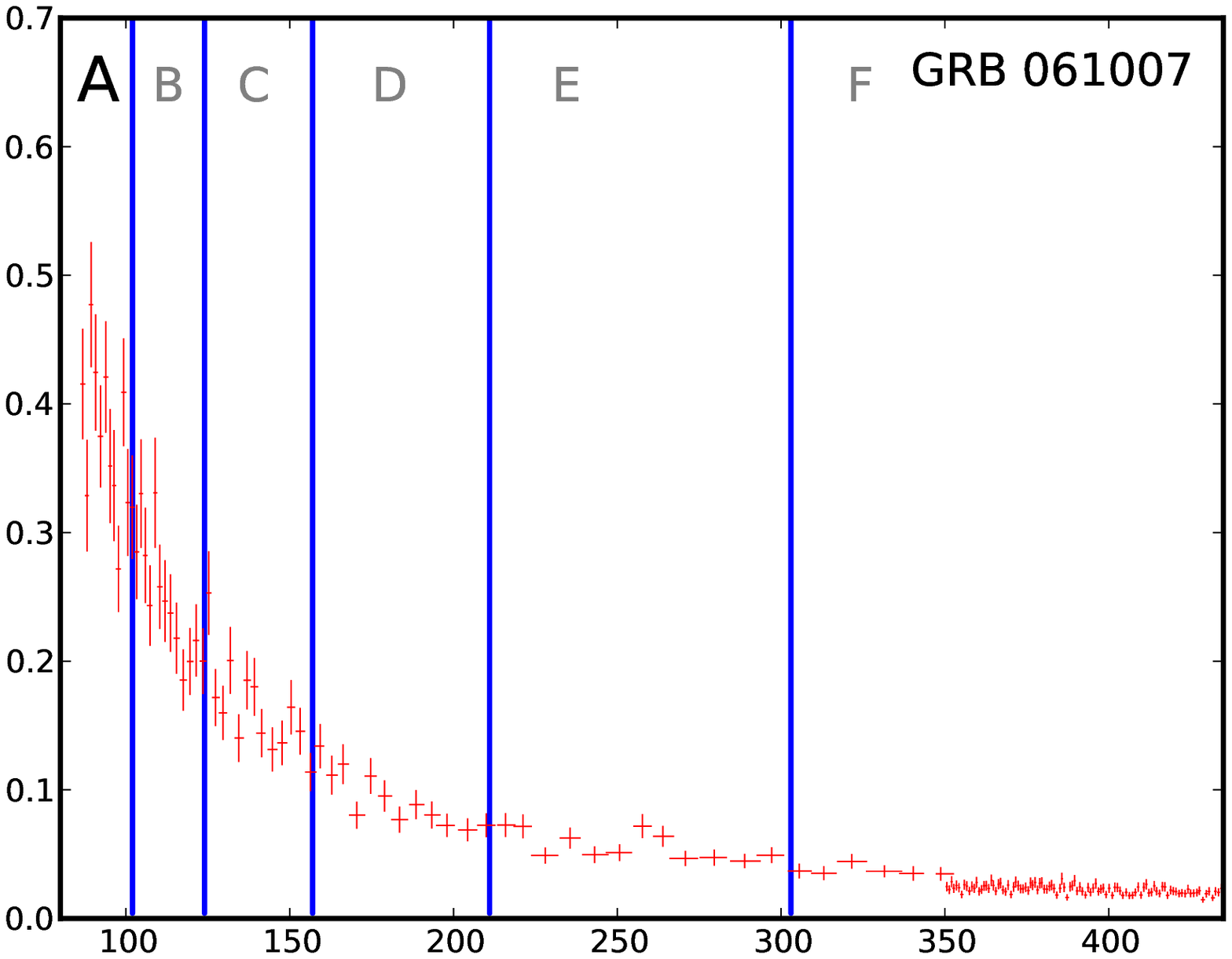}
\includegraphics[bb=18 210 575 609, width=0.475\textwidth]{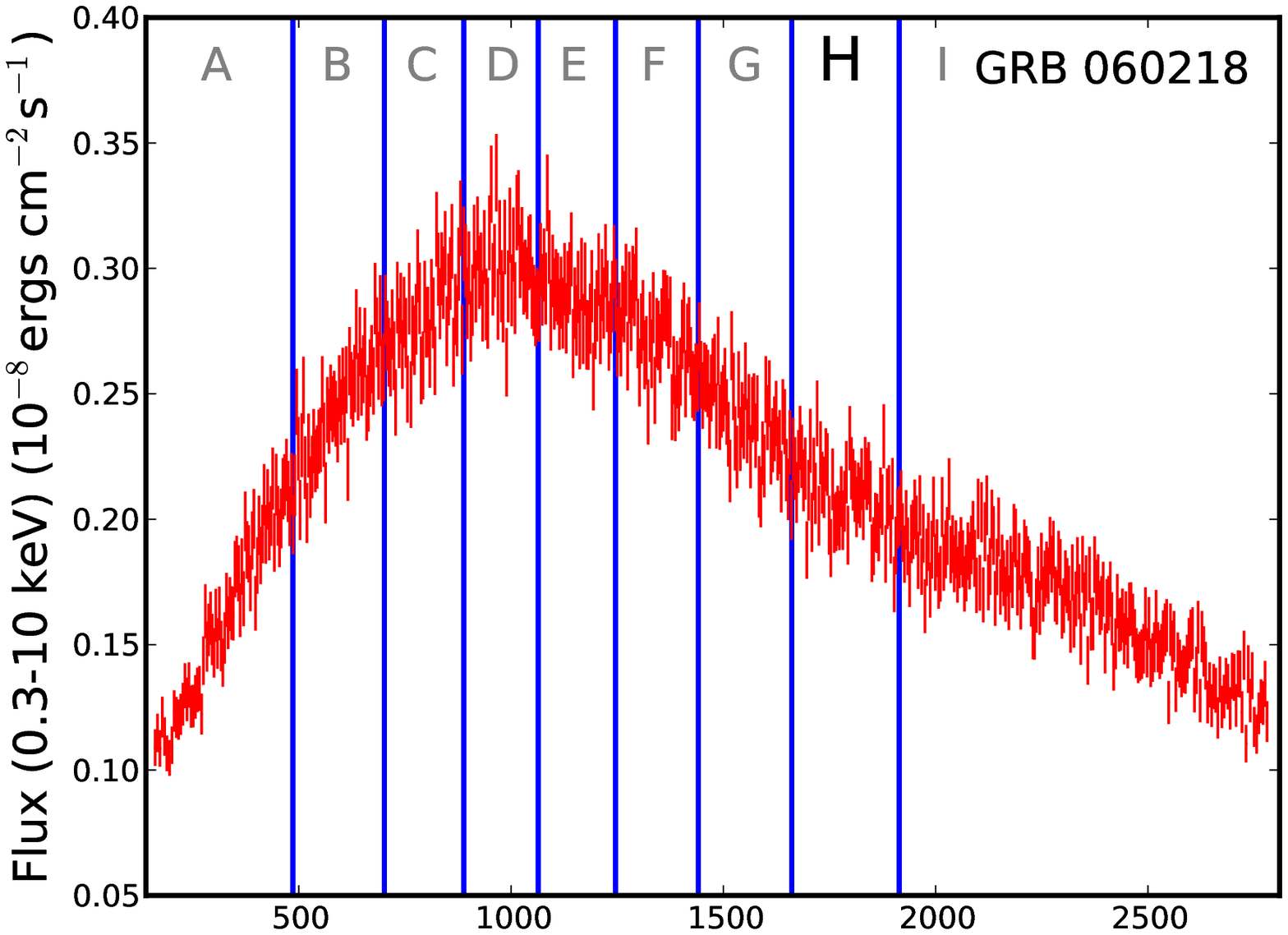}
\includegraphics[bb=18 210 575 609, width=0.475\textwidth]{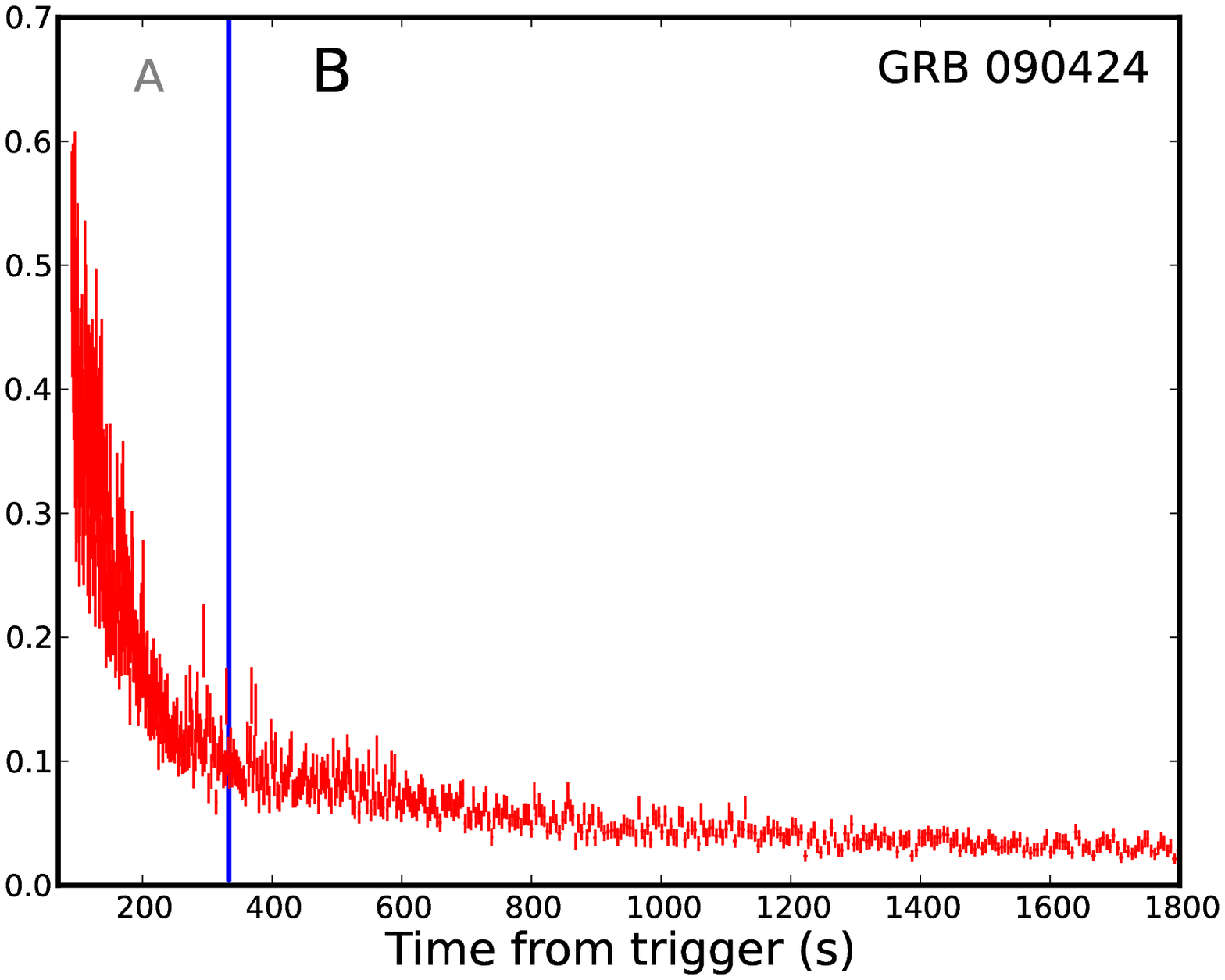}
\includegraphics[bb=18 196 575 609, width=0.475\textwidth]{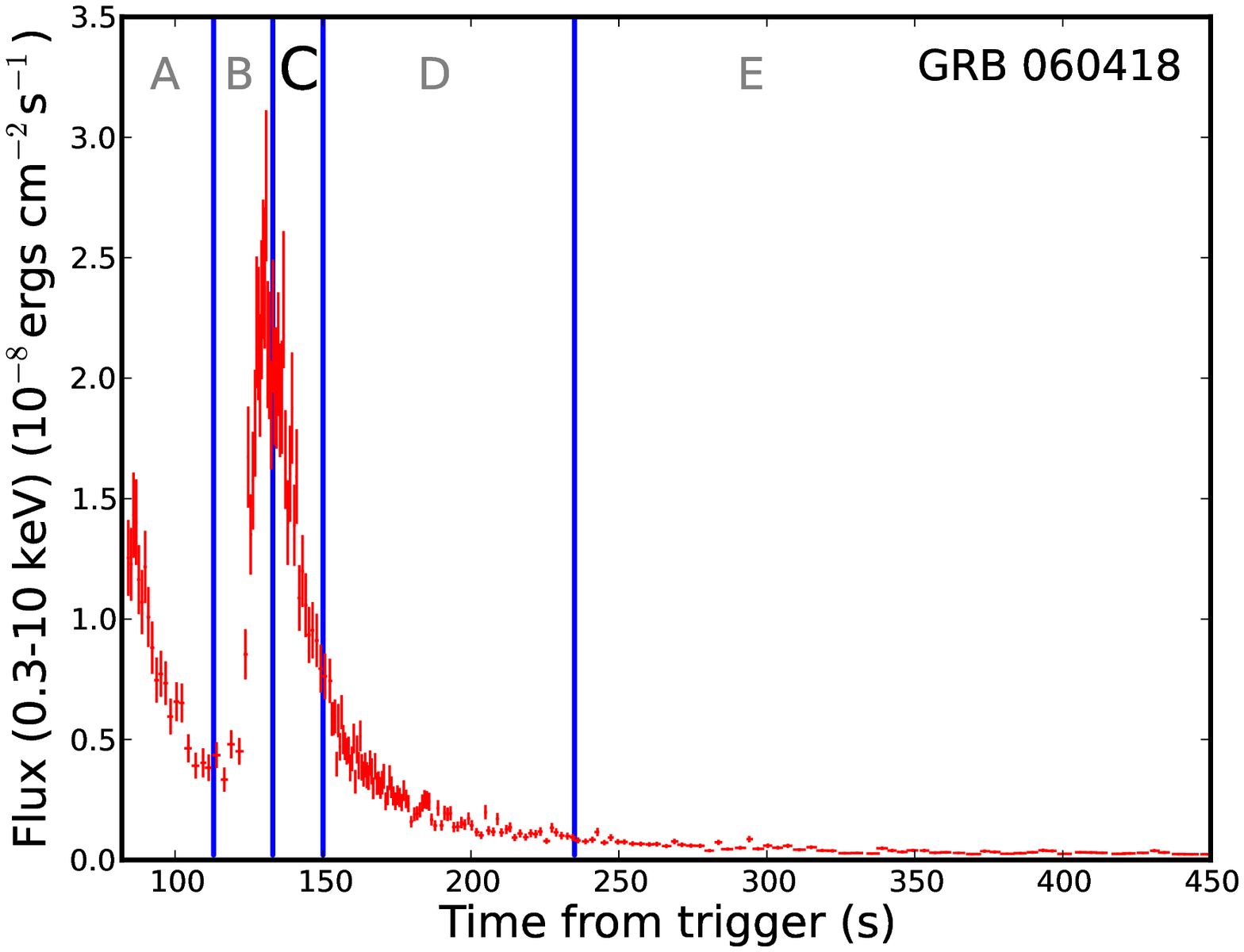}
\includegraphics[bb=-35 196 520 609, width=0.475\textwidth]{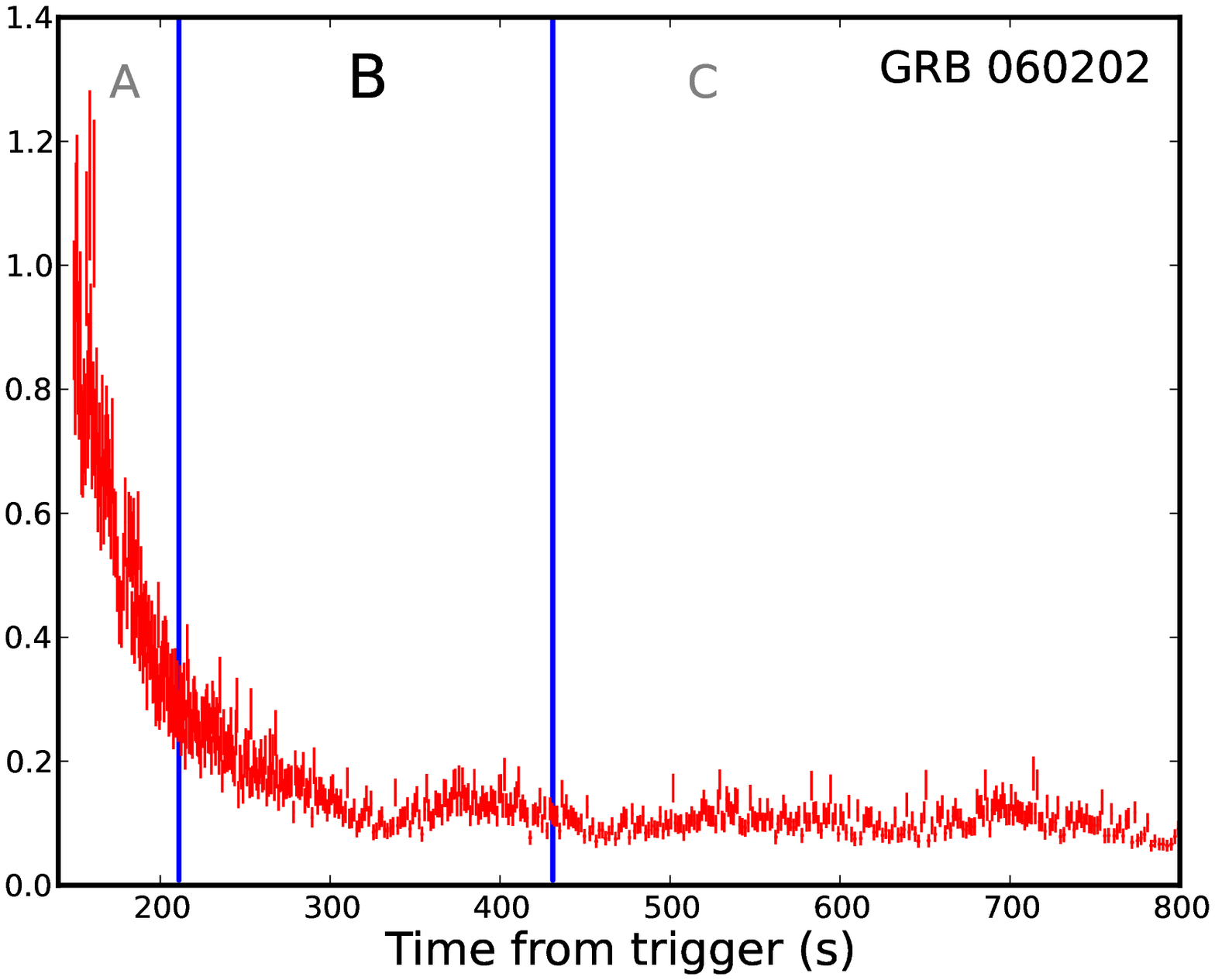}
\caption{Light curves for bursts with apparent thermal emission. The lines indicate where the data has been split into different spectra. Black letters indicate the time series with highest $\Delta\chi^2$.}
 \label{fig:grb061121_lc}
\end{figure*}

In this paper, we have processed data from \emph{Swift} BAT and XRT using the analysis tools within HEASOFT. 
BAT data have been included when available and the spectrum extracted using standard procedure, binning the data in the standard energy binning.
The WT mode data have been used from the XRT observations. This has been divided into time periods with
a minimum of 10\,000 counts for each
spectrum, except for GRB\,060418 where 5\,000 counts per spectrum was used.
The XRT light-curves for bursts with a detection of a thermal component are
presented in Fig.~\ref{fig:grb061121_lc} and the time periods delineated.
This way between two and nine spectra have been extracted for each burst. For data reduction the FTOOLs \emph{Swift}-specific sub package "xrtproducts" has been used. As centre-position the output from running "xrtcentroid" on the Photon Counting (PC) mode data has been used. The response files are from the \emph{Swift} repository. The spectra have been pile-up corrected following \cite{2006A&A...456..917R}.

We should note here that this sample of the brightest early-phase
\emph{Swift}-XRT bursts only overlaps with the six candidate bursts with
possible blackbody emission of \citet{2012MNRAS.427.2965S} in GRB\,100621A
as well as GRB\,090618 and GRB\,060218 which were claimed elsewhere.  The
rest of their bursts were not bright enough to enter our sample.  Of the
bursts we claim as possible detections in Table~\ref{tab:best}, the
above-mentioned bursts are detected in common. We also find detections in
GRB\,060202, 060418, 061007, 061121, and 090424. All of these bursts,
except GRB\,061007 are noted as initial possible candidates by
\citet{2012MNRAS.427.2965S}.

%
%
\section{Results}\label{results}

The extracted spectra were fit in \texttt{Xspec} with a Band model
\citep{1993ApJ...413..281B} with photoelectric absorption from both the
Milky Way ($z=0$, fixed) and the GRB host galaxy (variable and at the
redshift of the host).  This model was compared to a similarly absorbed
Band\,+\,blackbody model.  The Galactic foreground column densities were
fixed to values from the Leiden/Argentine/Bonn (LAB) Survey of Galactic
\ion{H}{1} \citep{2005A&A...440..775K} throughout the analysis.  The
equivalent hydrogen column density based on the dust extinction column would
be somewhat different, usually slightly lower \citep{2011A&A...533A..16W}. 
Using this value would typically increase the column density inferred at the
host galaxy redshift ($N_{\rm H}$), but would not affect the other fit
parameters substantially.  All fits have been done in two steps to reduce
the computational cost to a manageable level.  First the spectra for a given
burst were fit simultaneously to determine $N_{H}$ for the host galaxy.  The
absorption was then frozen to this value during individual fitting of the
spectra.  $N_{H}$ for the host was determined separately for the Band and
the Band\,+\,blackbody models.  The optical spectroscopic redshifts of the
GRBs were used to calculate model parameters in the host galaxy rest frame. 
Fit statistics and parameters of the best-fit blackbody for the time series
with the most significant blackbody detection for each burst in the sample
can be seen in Table~\ref{tab:best}.  The temperature and the
luminosity\footnote{The luminosity is given by the blackbody normalisation
in the xspec model: \[norm_{bb} =\frac{L_{39}}{D^2_{10}} =
\frac{L_{39}}{D_{L(10)}^2\,(1+z)^2}\] Here $L_{39}$ is the luminosity in
units of 10$^{39}$\,ergs/s, and $D_{10}/D_{L(10)}$ is the proper
motion/luminosity distance to the source in units of 10\,kpc.} were determined
from the fits.  Using these parameters and the Stefan-Boltzmann eq., the
apparent radiative surface area was determined, assuming a simple,
non-relativistic blackbody.  Under spherical geometry, the radius would then
be: $L = \sigma\,A\,T^4$, where $A = 4 \pi\,R^2$.


\begin{table*}
\caption{Blackbody parameters for the best fit time series for each burst in the sample. \\ Parameters have only been included if the improvement for the addition of a blackbody is better than $\Delta \chi^2$\,=\,25.}
\renewcommand*{\arraystretch}{1.7}
\begin{tabular}{@{} p{1.6cm}  p{1cm} p{0.9cm} p{0.9cm} l p{1.1cm} p{1.7cm} p{1.6cm}  p{1.3cm} c c c c c c c c @{}}
\hline\hline
GRB & Redshift & refs. & $\Delta \chi^2$ & bb lum.$^a$  & bb \%$^b$ & kT/keV & R$_{phot}$$^c$ & $\gamma$ & Time$^d$ & Time Series$^e$ \\ 
\hline

060202 & 0.783 & (1) & 28.1 & $2.7^{+0.1}_{-0.2}$ & 13 &  $0.38^{+0.01}_{-0.02}$ & $5.9^{5.1}_{5.2}$ & $60^{+14}_{-18}$ & 211--431 & B \\ 

060218 & 0.0331 & (2) & 73.4 & $0.0116^{+0.0007}_{-0.0006}$ & 0.24 & $0.156\pm{0.004}$ & $0.46^{+0.45}_{-0.46}$ & $40^{+9}_{-12}$ & 1661--1914 & H \\ 

060418 & 1.489 & (3) & 33.9 & $1.6^{+9}_{-0.7}$ & 3.5 & $0.53\pm{0.02}$ & $2.2^{+7.7}_{-1.2}$ & $<$254 & 150--235 & D \\ 

061007 & 1.262 & (4) & 36.1 & $119^{+12}_{-11}$ & 10 & $3.2^{+0.4}_{-0.3}$ & $1.9^{+0.17}_{-0.18}$ & $328^{+100}_{-64}$ & 86--102 & A \\ 

061121 & 1.314 & (4) & 49.4 & $257\pm{26}$ & 0.74 & $2.9\pm{0.2}$ & $6.7\pm{1.6}$ & $669^{+14}_{-18}$ & 72--86 & B \\ 

090424 & 0.544 & (5) & 49.8 & $0.16^{+0.01}_{-0.04}$ & 27 & $0.228\pm{0.006}$ & $19^{+23}_{-17}$ & $26^{+8}_{-9}$ & 333--5554 & B \\ 

090618 & 0.54 & (6) & 46.9 & $1.81^{+0.08}_{-0.08}$ & 17 & $0.74^{+0.08}_{-0.06}$ & $12\pm{2}$ & $<$1058 & 138--145 & C \\ 

100621A & 0.542 & (7) & 36.5 & $0.73^{+0.07}_{-0.09}$ & 23 & $0.38^{+0.39}_{-0.36}$ & $2.8^{+2.6}_{-2.6}$ & $40^{+11}_{-10}$ & 141--40540 & C \\ 

\hline

060124 & 2.300 & (4) & 11.9 & $<$25 & $<$0.25 & --- & --- & --- & 557--635 & B \\ 

060510B & 4.9 & (8) & 9.26 & $<$136 & $<$38 & --- & --- & --- & 127--252 & A \\ 

060526 & 3.221 & (9) & 8.17 & $<$8.8 & $<$4.1 & --- & --- & --- & 81.5--282 & A \\ 

060614 & 0.125 & (10) & 14.2 & $<$0.14 & $<$15 & --- & --- & --- & 97--109 & A \\ 

060729 & 0.543 & (4) & 11.6 & $<$37 & $<$23 & --- & --- & --- & 130--146 & A \\ 

060814 & 1.92 & (11) & 24.0 & $<$0.25 & $<$52 & --- & --- & --- & 168--75055 & C \\ 

060904B & 0.703 & (4) & 15.5 & $<$1.8 & $<$33 & --- & --- & --- & 77--178 & A \\ 

071031 & 2.692 & (4) & 19.3 & $<$70 & $<$22 & --- & --- & --- & 109--178 & A \\ 

080310 & 2.42 & (12) & 14.5 & $<$8.6 & $<$3.1 & --- & --- & --- & 215--287 & B \\ 

080319B & 0.937 & (13) & 20.1 & $<$7.6 & $<$0.05 & --- & --- & --- & 90--120 & B \\ 

080607 & 3.036 & (14) & 7.45 & $<$222 &$<$18 & --- & --- & --- & 123--144 & B\\ 

080928 & 1.692 & (4) & 19.1 & $<$22 & $<$8.0 & --- & --- & --- & 210--248 & B\\ 

081008 & 1.9685 & (15) & 14.7 & $<$35810 & $<$100 & --- & --- & --- & 94--144 & A \\ 

081028 & 3.038 & (16) & 8.60 & $<$3.5 & $<$14 & --- & --- & --- & 341--61920 & B \\ 

090417B & 0.345 & (17) & 2.87 & $<$0.1 & $<$1.8 & --- & --- & --- & 702--1497 & B \\ 

090516A & 4.109 & (18) & 14.6 & $<$325 & $<$46 & --- & --- & --- & 171--136 & A \\ 

090715B & 3.0 & (19) & 9.33 & $<$82 & $<$7.7 & --- & --- & --- & 219--274 & C \\ 

100814A & 1.44 & (20) & 9.17 & $<$41 & $<$3.6 & --- & --- & --- & 94--157 & A \\ 

110205A & 2.22 & (21) & 12.0 & $<$76 & $<$7.1 & --- & --- & --- & 273--325 & D \\ 

110801A & 1.858 & (22) & 18.5 & $<$57 & $<$13 & --- & --- & --- & 380--424 & C \\ 

\hline
\end{tabular}
\\
$^a$ Blackbody luminosity in $10^{48}$\,ergs\,s$^{-1}$ \\
$^b$ Percent of total luminosity in the thermal component. \\
$^c$ Photospheric radius in $10^{13}$\,cm \\
$^d$ Seconds since BAT trigger \\
$^e$ Part of light curve that contains best fit improvement \\
References: (1) \cite{2007ApJ...656.1001B}; (2) \cite{2006GCN..4792....1M}; (3) \cite{2006ApJ...648L..93P}; (4) \cite{2009ApJS..185..526F}; (5) \cite{2009GCN..9243....1C}; (6) \cite{2009GCN..9518....1C}; (7) \cite{2010GCN..10876...1M}; (8)\cite{2006GCN..5104....1P}; (9) \cite{Jakobsson2006}; (10) \cite{2006Natur.444.1050D}; (11) \cite{2012ApJ...752...62J}; (12) \cite{2012A&A...545A..64D}; (13) \cite{2008Natur.455..183R}; (14) \cite{2009ApJ...691L..27P}; (15) \cite{2008GCN..8350....1D}; (16) \cite{2008GCN..8434....1B}; (17) \cite{2009GCN..9156....1B}; (18) \cite{2009GCN..9383....1D}; (19) \cite{GCN9673};   (20) \cite{2010GCN..11089...1O}; (21) \cite{2011GCN..11638...1C}; (22) \cite{2011GCN..12234...1C}
\label{tab:best}
\end{table*}

In Table~\ref{tab:best}, the $\Delta\chi^2$ values are given for the two models. The actual probabilities as seen in Table~\ref{tab:fit} are inferred from a Monte Carlo analysis, and gives the likelihood that the added blackbody is just a better fit per random chance. As can be seen from Table~\ref{tab:best} the spectral fit seems generally to be improved with the addition of a blackbody. The blackbody model only adds two free parameters, so the fact that $\chi^2$ \textit{(was wrong before, should not have been $\Delta\chi^2$)} decrease by more than a factor of two in all but one burst indicates that an extra component is needed in addition to the Band function. 


To test the statistical significance of the fit improvement, we used a Monte
Carlo method, generating 10\,000 artificial spectra from the single Band
function model.  We did this for every burst with a large improvement in
$\chi^2$ ($\Delta\chi^2\geq25$, the 3\,$\sigma$ limit from the first three
time series analysed), since it is extremely unlikely, that a small
$\Delta\chi^2$ value will turn out to be significant, and because the Monte
Carlo analysis is computationally expensive.  The distribution of $\chi^2$
was found to be the same for different time series in the same burst, but
not compatible between different bursts using the Kolmogorov-Smirnoff test.  For
bursts with no match in $\Delta\chi^2$ in the first 10\,000 spectra, a
further 10\,000 were done to improve statistics.  The detection probabilites
for an additional blackbody component resulting from these simulations are
provided in Table~\ref{tab:fit} for each time series. We adopt a thermal component
detection criterion of less than one in 10\,000 over the null hypothesis in
at least one of the time series. This corresponds to a detection probability
of better than $\sim4\sigma$.

Early afterglow thermal emission has already been reported for GRB\,060218
\citep{2006Natur.442.1008C}, and extensively analysed
\citep{2007ApJ...667..351W, 2007MNRAS.375..240L, 2007MNRAS.375L..36G, 2007MNRAS.382L..77G, 2006A&A...454..503S}.  The detection of thermal emission in
GRB\,090618 \citep{2011MNRAS.416.2078P} and GRB\,100621A
\citep{2012MNRAS.427.2965S} has also been reported, though not as
extensively, and so we include our analysis of these bursts here for
comparison.  In addition to these bursts we report the detection of thermal
emission in GRBs 060202, 060418, 061007, 061121, and 090424.

\subsection{GRB\,061007}\label{grb061007}

The spectral fitting for GRB\,061007 is improved for the added blackbody for
time series A to $>4\sigma$ confidence (that is, the Monte Carlo analysis
found no $\Delta\chi^2$ match in 20\,000 simulated spectra).  We also
performed 10\,000 simulations separately for time series B, where 0.16\%
(approximately $3\sigma$ confidence) of the faked data had a $\Delta\chi^2$
as large or larger than the real one.  The component is very bright,
accounting for respectively 10 and 11\% of the total luminosity in series A
and B.  With luminosities of up to $1.9\times10^{50}$\,ergs\,s$^{-1}$ as
well as blackbody temperatures of several keV, this thermal emission is both
more luminous and hotter than those previously reported
\citep{2006Natur.442.1008C,2011MNRAS.416.2078P,2012MNRAS.427.2965S}.  The
spectra are plotted in Fig.~\ref{fig:spectra}, top panel.



\subsection{GRB\,090424}

The thermal emission from GRB\,090424 looks more like the ``classical"
thermal component, with low luminosity and temperatures of $\sim0.2$\,keV,
similar to, for instance, GRB\,060218.  We performed the Monte Carlo
analysis for both series separately.  For time series B no $\Delta\chi^2$
match was found in 20\,000 spectra.  For the time series A about 0.2\%
($>3\sigma$) match the real value.  Compared to the burst's low total
luminosity, the thermal emission is very bright, constituting about one
fourth of the total luminosity.  The spectra can be seen in
Fig.~\ref{fig:spectra}, second panel.



\subsection{GRB\,061121, 060202 and 060418}\label{rest}

\begin{figure*}
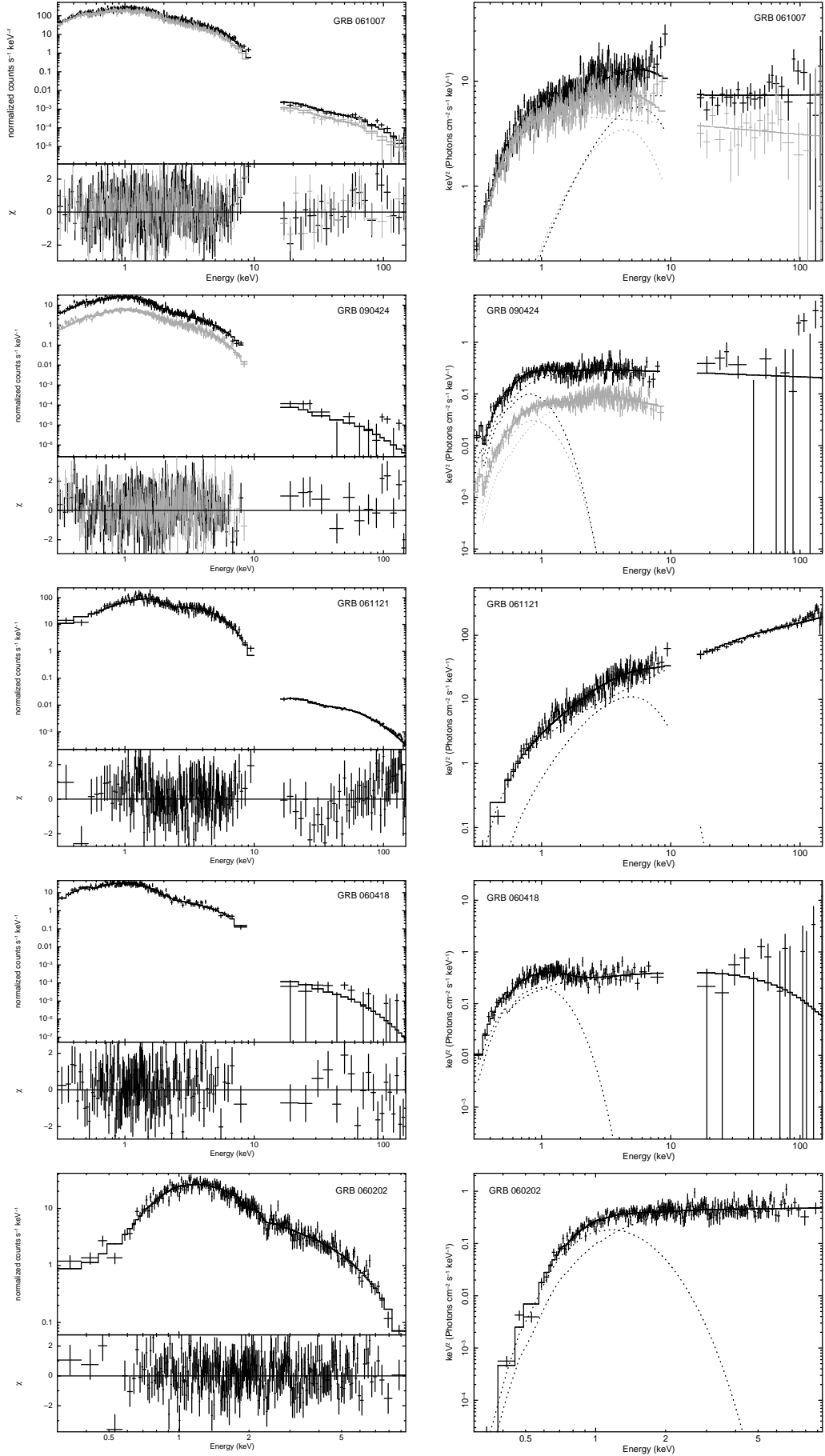

\centering
\includegraphics[bb=70 16 575 732, angle=-90, width=0.38\textwidth]{GRB061007_bb_all.ps}
\includegraphics[bb=70 16 575 732, angle=-90, width=0.38\textwidth]{GRB061007_unf.ps}
\includegraphics[bb=70 16 575 732, angle=-90, width=0.38\textwidth]{GRB090424_bb.ps}
\includegraphics[bb=70 16 575 732, angle=-90, width=0.38\textwidth]{GRB090424_unf.ps}
\includegraphics[bb=70 16 575 732, angle=-90, width=0.38\textwidth]{GRB061121_bb.ps}
\includegraphics[bb=70 16 575 732, angle=-90, width=0.38\textwidth]{GRB061121_unf.ps}
\includegraphics[bb=70 16 575 732, angle=-90, width=0.38\textwidth]{GRB060418_bb.ps}
\includegraphics[bb=70 16 575 732, angle=-90, width=0.38\textwidth]{GRB060418_unf.ps}
\includegraphics[bb=70 16 575 732, angle=-90, width=0.38\textwidth]{GRB060202_bb.ps}
\includegraphics[bb=70 16 575 732, angle=-90, width=0.38\textwidth]{GRB060202_unf.ps}
\caption{Spectra for GRBs 061007 (time series A and B), 090424 (time series A and B), 061121 (time series B), 060418 (time series C) and 060202 (time series B) respectively. Left panel shows the spectra including the residuals, right panel shows the unfolded spectra. For GRBs 061007 and 090424 black lines show time series A, while grey lines show time series B.}
 \label{fig:spectra}
\end{figure*}

GRBs 061121, 060202 and 060418 are only good candidates for thermal emission
in one time series.  Spectra can be seen in Fig.~\ref{fig:spectra}. 
We note that the parameters for GRB\,060418 are not very well constrained.  All
parameters for the detections are similar, with temperatures around 1\,keV,
luminosities corresponding to few percent of the total, and apparent
host-frame blackbody radii around $10^{11}$--$10^{12}$\,cm.
The general tendency seems to be for the thermal component to 
be detected in the declining phase after a flare
(though for GRB\,061121 the time series includes the peak of the flare), see Fig~\ref{fig:grb061121_lc}. 

\subsection{Previous detections}

Our statistical analysis is relatively conservative as we require a
$>4\sigma$ detection.  Furthermore, we use a Band model as our underlying
continuum, which allows for a certain curvature in the spectra without
invoking an additional component.  For example for a burst such as
GRB\,090618 where thermal emission has been claimed previously, we only have
a clear detection by our criteria in the very late phase.  For our
claimed detections, only GRB\,061007 is not noted to possibly have a thermal component (i.e.\ to have a significant deviation from a single power-law) in the analysis of
\citet{2012MNRAS.427.2965S}.  However, this may not be surprising, since we
find a highly significant detection only in the first epoch and the signal
may be washed out over the rest of the burst.  GRBs 060202, 060418, 061121,
and 090424 are all flagged as initial possible candidates in
\citet{2012MNRAS.427.2965S}, but are excluded from their final list for a variety of reasons. We examine these reasons below and discuss why these GRBs have thermal components.  GRBs\,060202 and 061121 were excluded because the
significance of their detections was too sensitively dependent on the column
density used in the fits.  For our analysis, we do not find that varying our
column density, which is based on a simultaneous fit of all the data, has a
notable effect on the statistical significance of the detection for these
bursts. Within a 90\,\% c.l $\Delta\chi^2$ varies with a maximum of 3, while the temperature 
for the blackbody fit remains constant within errors. 
Furthermore it was noted that GRB\,061121 had a high temperature and redshift.  We do not include the
temperature or redshift attributes of the thermal components in our search
criteria.  GRB\,060418 was excluded by \citet{2012MNRAS.427.2965S} because
of strong flaring activity in the light curve.  We do not exclude flaring
light curves in our analysis a priori since we fit the data with a Band model
as the underlying continuum and include the BAT data in our analysis where
it is available.  Finally, GRB\,090424 was found to have a low statistical
significance in their time-sliced spectra.  Our detection comes from a
spectrum over a much wider time range (333--5554\,s compared to
700--1000\,s) and hence has a far higher signal-to-noise ratio.

\begin{figure}
\includegraphics[width=\columnwidth]{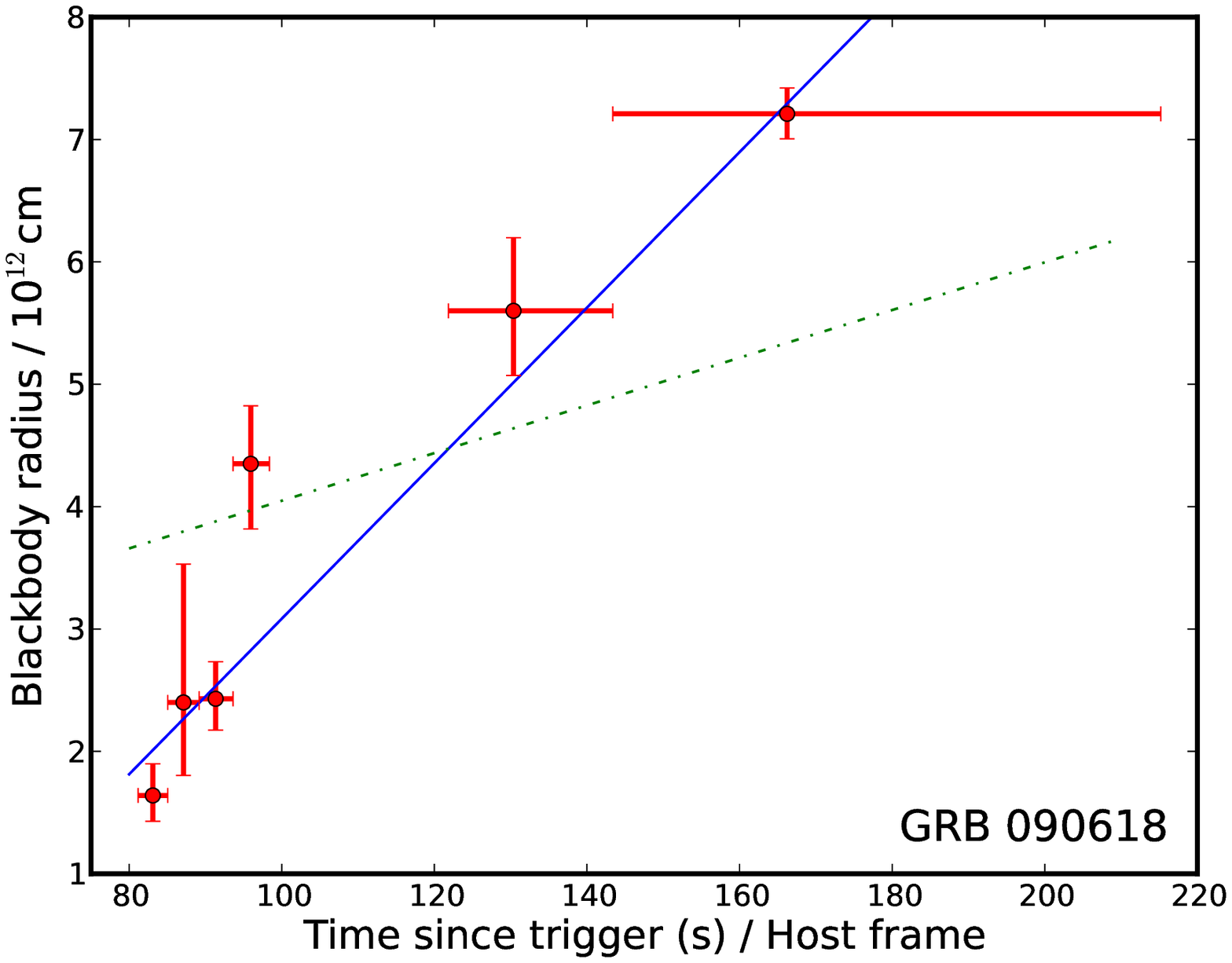}
\caption{Expansion of the blackbody radius for GRB\,090618. The solid line shows the best fit for a constant velocity ($v=3.5\,c$. For comparison a fit with an apparent velocity equal to the speed of light ($v=c$, dot-dashed) is also shown, but does not fit the data well.}
\label{fig:hostframe090618}
\end{figure}

We track the evolution of the temperature and luminosity of GRB\,090618 and
find, in the host galaxy rest-frame an apparent expansion with a best-fit
speed of $10.5^{+9.5}_{-2.1}\times10^{10}$\,cm\,s$^{-1}$ (see Fig.~\ref{fig:hostframe090618}).  This is $3.5^{+3.0}_{-0.7}$ times the speed of light.  We therefore conclude that the thermal component in
GRB\,090618 must be treated relativistically. \\
\\

\begin{table}
\caption{$\Delta\chi^2$'s and probabilities for all time series of bursts where at least one series has $\Delta \chi^2$\,$\geq$\,25. (GRB\,060218 has been omitted).}
\renewcommand*{\arraystretch}{1.5}
\begin{tabular}{@{} p{1.1cm} p{1.1cm} p{1.7cm} p{1.3cm} l c c c @{}}
\hline\hline
Time$^{a}$ & Nr.$^{b}$  & Band $\chi^2$/d.o.f & $\Delta\chi^2$ & Probability \\ \hline
\hline \multicolumn{4}{ l }{{GRB\,060202 ($z\,=\,0.783$)}} \\ \hline
171.3 & A & 361.29/329 & 6.7 & 0.12 \\ 
295.3 & B & 342.65/268 & 28 &  0.000068\\ 
684.1 & C & 370.49/342 & 6.4 & 0.13 \\ 
\hline \multicolumn{4}{ l }{{GRB\,060418 ($z\,=\,1.489$)}}  \\ \hline 
 93.00 & A & 174.59/192 & 5.0 & 0.16 \\ 
 128.5 & B & 245.48/202 & 2.1 & 0.50 \\ 
 140.2 & C & 216.15/201 & 13 & 0.0078 \\ 
 171.9 & D & 240.63/211 & 34 & $<5\times10^{-5}$ \\ 
 542.1 & E & 157.76/176 & 6.9 & 0.076 \\ 
\hline \multicolumn{4}{ l }{{GRB\,061007 ($z\,=\,1.261$)}} \\ \hline
94.22 & A & 340.71/316 & 36 &  $<3.3\times10^{-5}$ \\ 
113.1 & B & 336.81/308 & 27 &  0.0016 \\ 
139.5 & C & 379.37/319 & 11 & 0.10 \\ 
181.1 & D & 356.18/317 & 4.0 & 0.56 \\ 
251.8 & E & 330.56/316 & 8.1 & 0.21 \\ 
372.8 & F & 330.56/315 & 12 & 0.070 \\ 
612.8 & G & 387.64/316 & 14 & 0.041 \\ 
1286  & H & 312.30/264 & 24 & 0.0033 \\ 
\hline \multicolumn{4}{ l }{{GRB\,061121 ($z\,=\,1.314$)}} \\ \hline
68.80 & A & 298.45/323 & 2.0 & 0.61 \\ 
78.72 & B & 456.34/339 & 49 &  $<5\times10^{-5}$ \\ 
102.6 & C & 257.15/235 & 4.1 & 0.27 \\ 
371.3 & D & 228.59/194 & 20 & 0.0003 \\ 
\hline \multicolumn{4}{ l }{{GRB\,090424 ($z\,=\,0.544$)}} \\ \hline
165.0 & A & 387.76/305 & 21 &  0.0020 \\ 
618.1 & B & 380.28/328 & 50 & $<3.3\times10^{-5}$ \\ 
\hline \multicolumn{4}{ l }{{GRB\,090618 ($z\,=\,0.54$)}} \\ \hline 
 128.1 & A & 191.41/208 & 11 & 0.034 \\ 
 134.3 & B & 244.24/226 & 19 & 0.0045 \\ 
 140.8 & C & 255.60/241 & 27 &  0.00095 \\ 
 147.8 & D & 254.21/231 & 16 & 0.0074 \\ 
 155.6 & E & 234.05/214 & 5.7 & 0.21 \\ 
 165.1 & F & 214.43/208 & 6.6 & 0.16 \\ 
 178.3 & G & 200.51/185 & 1.3 & 0.83 \\ 
 200.9 & H & 253.39/233 & 29 & 0.00057 \\ 
 256.1 & I & 368.69/300 & 47 &$<3.3\times10^{-5}$ \\ 
\hline \multicolumn{4}{ l }{{GRB\,100621A ($z\,=\,0.542$)}} \\ \hline
88.05 & A & 249.80/311 & 7.6 & 0.041 \\ 
117.3 & B & 322.56/286 & 8.8 & 0.028 \\ 
173.3 & C & 257.35/212 & 37 & $<5\times10^{-5}$  \\ 
 \hline
\end{tabular}
\\$^{a}$ Mean time in seconds after BAT trigger time.
\\$^{b}$ Time series
\label{tab:fit}
\end{table}

%
%

\section{Origin of the thermal component}\label{discussion}
\subsection{SN shock breakout}

It has been suggested that the thermal emission found in previous GRB early
afterglows is due to a SN shock break-out from the stellar winds surrounding
the progenitor.  While this may be a plausible model for low-luminosity
systems, it seems implausible that the very large luminosities discovered
here could possibly be related to a shock breakout from a SN.  Models of SN
shock breakouts confirm this \citep[e.g.][]{2007MNRAS.375..240L}. Typical 
values reported are $10^{47}$\,ergs for the break-out energy and a
blackbody temperature of $1$\,keV. This energy is lower than any we observe in 
the soft X-ray thermal component. Even attempts to explain GRB\,060218's thermal 
component, (the lowest blackbody luminosity in our sample) with an asymmetric 
explosion \citep{2007ApJ...667..351W} have been shown by 
\cite{2007MNRAS.382L..77G} to require a deal of fine tuning. The fact that we obtain very large luminosities for 
both GRB\,061007 and GRB\,061121 (about four orders of magnitude larger than 
GRB060218), but that the properties of the extra component are not qualitatively 
dissimilar to previous thermal components found, suggests strongly that not only is 
a SN shock break-out not the origin in these cases, but it may not be in
most other cases where this has been discovered either. \cite{2008Natur.453..469S} reported the case of possible shock breakout in a SN without an accompanying GRB, SN2008D. It has also been suggested that this was photospheric emission from a mildly relativistic jet \citep{2008Sci...321.1185M}. The case has been throughly studied, \citep{2011ApJ...726...99V,2010A&A...522A..14G,2009ApJ...700.1680T}, but without conclusive results. With a total energy of $E_X\approx2\times10^{46}$\,ergs, this burst is consistent with what could be expected for a break-out, and with the limited photon statistics we cannot distinguish the origin of the emission using the spectra. For the GRBs though, we should look elsewhere for the origin of an apparently thermal component in the late
prompt/early afterglow soft X-ray emission.

\subsection{Alternative models}

While much progress has been made because of the afterglows of GRBs, the
origin of, and mechanism behind the prompt phase and early afterglow are
still uncertain.  Emission from a cocoon surrounding the
jet has been proposed \citep[e.g.][]{2013ApJ...764L..12S}, but the model does not 
seem to explain the energies and expansion velocities reported here. 

The models traditionally used to model the high energy
spectra, a cut-off power-law or smoothly broken power-law
\citep{1993ApJ...413..281B}, are empirical and not strongly motivated from a
physical understanding of the emission process.  The main part of the
radiation is often considered to be non-thermal, with high energy photons
originating from synchrotron and/or inverse Compton processes in the
ultra-relativistic jet \citep[e.g.][]{1996ApJ...466..768T,1997ApJ...488..330C,2011A&A...526A.110D}). 
However, based on the detection of components in the prompt gamma-ray
emission with a blackbody-like spectral shape in addition to a power law
\citep{2005ApJ...625L..95R}, and on difficulties in reproducing low energy
spectral indices with synchrotron models \citep{1998AIPC..428..359C}, a
trend has been growing to attribute much of the prompt-phase emission to the
photosphere at the head of the ultra-relativistic jet
\citep[e.g.][]{2007RSPTA.365.1171P,2010arXiv1003.2582P,2013arXiv1301.3920L}. 
Such emission emerges from an optically thick plasma, generally not in
thermal equilibrium, producing a quasi-thermal spectrum
\citep{2009ApJ...702.1211R}.  Observationally, this high energy photospheric
emission decays in luminosity and temperature as a power-law in time
\citep{2009ApJ...702.1211R}.  If that trend continues, it is not
unreasonable to suppose that it may appear at the end of the prompt phase in
soft X-rays as an apparently thermal component with high apparent luminosity
and temperature.  Under this hypothesis, we can then model the excess
component as late photospheric emission.

Using eqs.~5 and 7 found in \cite{2012MNRAS.420..468P} (modified to our parameters), we can calculate the Lorentz factor and the photospheric radius:
\[r_{ph} = R^{host}_{bb}\times\frac{\gamma}{\,\xi\,(1+z)^{2}}\] 
\[ \gamma = [(1+z)^2\,D^2_L\frac{F^{obs}_{bb}\sigma_T}{2\,m_p\,c^3\,R^{host}_{bb}}]^{1/4}\times(L_{tot}/L^{obs}_{bb})^{1/4}\] 
Taking the model luminosities found for each time series we get values for the Lorentz factors, as seen in Table~\ref{tab:best}. These values should be corrected with a factor expressing the ratio of the blackbody and total luminosity that is in the relevant burst epoch, as the equation uses the values over the entire burst, but as the luminosity ratio is to the power of 1/4, the correcting factor will be close to 1. We then use these values to calculate the photospheric radii (also shown in Table~\ref{tab:best}), where $\xi$ is a geometrical factor close to one (we set $\xi$\,=\,1 exactly).

\begin{figure}
\includegraphics[angle=0,bb=18 180 594 612,clip=,width=\columnwidth]{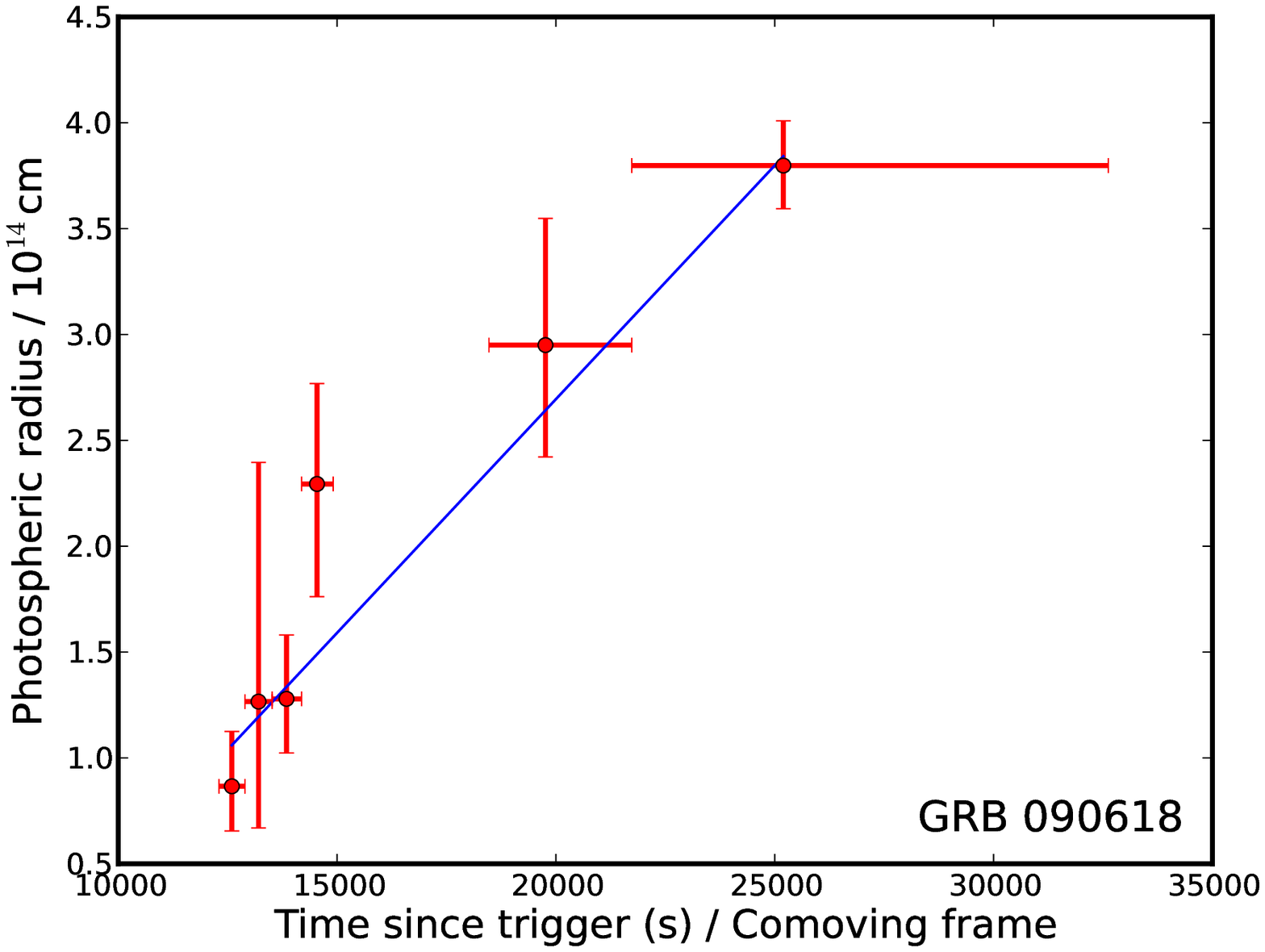}
\includegraphics[angle=0,bb=18 180 594 612,clip=,width=\columnwidth]{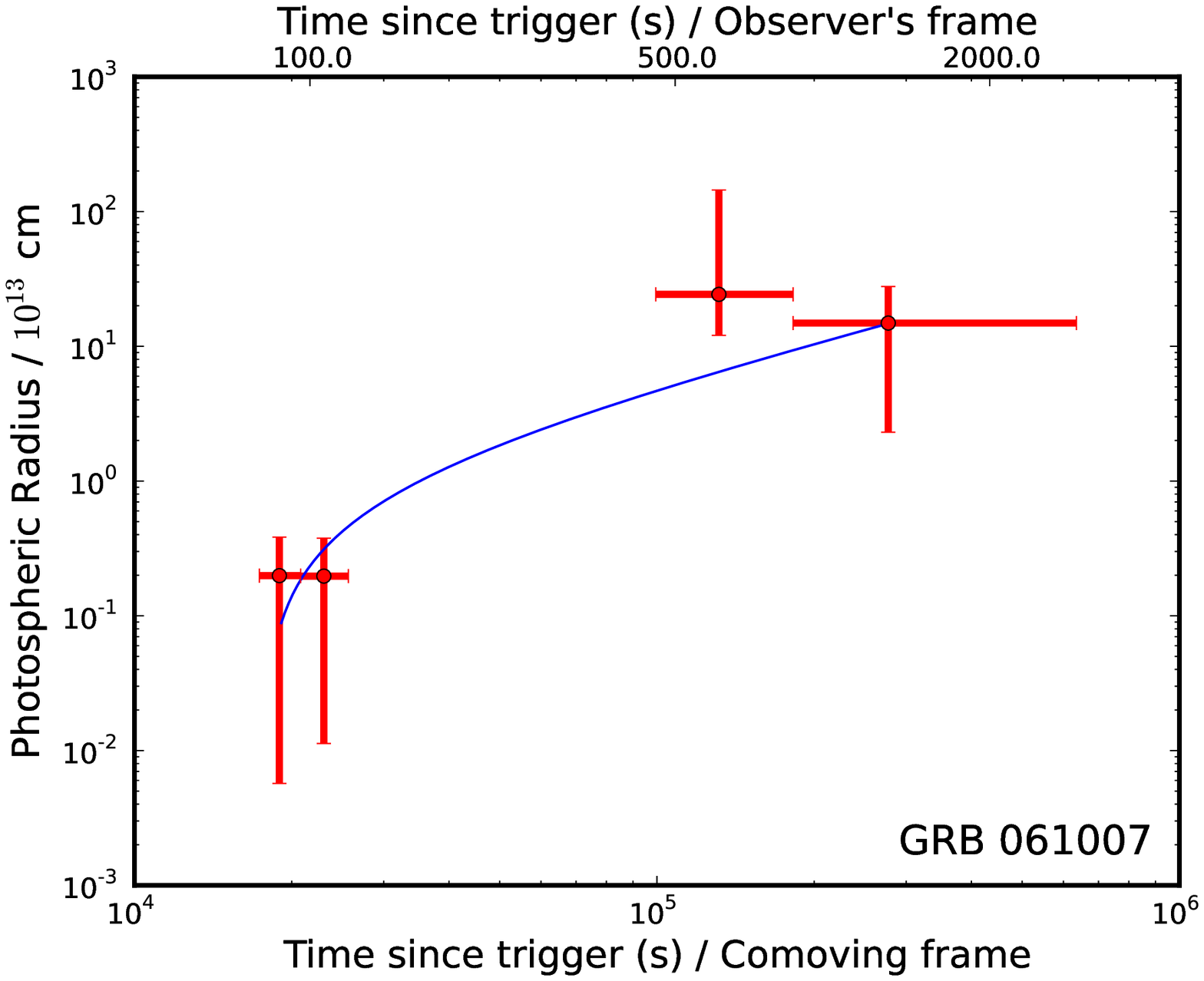}
\caption{Evolution of the photospheric radius for GRBs 090618 and 061007. The photospheric radii are derived from eqs. 5 and 7 in \cite{2012MNRAS.420..468P}. The solid lines show the best fit for constant velocity. Only the velocity of GRB\,090618 is consistent with expansion close to $c$ in the jetframe using the calculated asymptotic Lorentz factor for the burst, indicating a slowdown of the jet in the other bursts by the time of the XRT observations.}
\label{fig:radius}
\end{figure}

The Lorentz factors calculated here are asymptotic values, valid for the
coasting phase of the jet, so they should be similar to the values
calculated from the prompt phase emission.  As seen in Table~\ref{tab:best},
the Lorentz factors we get are consistent with the emission being late
photospheric, with values between ten and several hundreds.  Another
consistency check is apparent superluminal motion of the photospheric radii. 
In order to look for this, we fitted the photospheric expansion assuming this to be constant throughout.  
We checked all bursts with detections better than 2\,$\sigma$ in more than one time series, using these for the fit.  
Fig.~\ref{fig:radius} shows fits for GRBs 090618 and 061007.  


For GRB\,090618 we fit an apparent velocity of $111\,c$, which corresponds
to $0.6^{+0.4}_{-0.2}\,c$ in the jet frame using the calculated asymptotic
Lorentz factor, consistent with a relativistic expansion close to that
Lorentz factor.  For GRBs 061007 and 061121 the fitted non-relativistic
velocities are $93\,c$ and $68\,c$ respectively.  However, using the
asymptotic Lorentz factors in this case yields jetframe velocities only a
small fraction of the speed of light. We conclude that for these bursts the
jet has started to slow down and expand, so the assumption of a constant
Lorentz factor is no longer valid.  Our sample then includes all cases, both
the jet coasting phase, with velocities near the speed of light, as seen in
GRB\,090618, the deceleration phase as seen for GRBs 061007 and 061121, and
the case where the jet has slowed down completely, seen by the constant
radius (photospheric as well as simple blackbody) of GRBs 060418 and 090424.

In principle, the observed evolution of the thermal component in the soft
X-ray should be compatible with those observed at early times in the
gamma-ray regime.  While we have only one GRB in our sample which has a
reported prompt phase gamma-ray thermal component, GRB\,061007 (see below),
as reported by \cite{2009ApJ...702.1211R}, the evolution in temperature and
luminosity behaves consistently across bursts, with an initial increase in
value until a break occurs after a few seconds, and then a decay with a
power-law index between $-4.5$ and $-0.8$ for the luminosity and $-1.3$ and
$-0.3$ for the temperature.  Using the points we have from the bursts with
good fit over several time series to calculate power-law indices, we get
values largely consistent with these for GRB\,061007, 090618, 061121,
090424 and 060218; luminosity: $-0.2^{+0.2}_{-1.1}$, $-0.7^{+0.3}_{-1.9}$,
$-1.9^{+1.0}_{-8.5}$, $-0.8^{+0.6}_{-0.6}$, and $0.7^{+0.2}_{-0.6}$, and temperature: $-1.0^{+0.5}_{-1.0}$,
$-0.4^{+0.2}_{-1.4}$ and $-0.8^{+0.4}_{-4.9}$, $0.04^{+0.05}_{-0.04}$, and $-0.12^{+0.18}_{-0.03}$ respectively. 
As far as we are aware there is no theoretical reason why these power-law
decay trends necessarily have to continue indefinitely, but the fact that
our results for the luminosity and temperature decay rates lie in the range
observed from the gamma-ray prompt phase is encouraging for the model.

GRB\,061007 is of special interest, since \cite{2011MNRAS.414.2642L} find
the prompt emission of this burst to be dominated by apparent thermal
emission.  They find that the evolution in temperature and flux follow the
light curve.  By the time the XRT started observing, the light curve is in
constant decline, so the temperature and flux would be expected to decay, as
we find in our analysis of the XRT data.  We cannot compare decay indices,
as \citet{2011MNRAS.414.2642L} observe before the onset of the declining phase. 
The thermal component started off very strong, accounting for about 75\% of the total
luminosity.  By the time of the XRT observation it has fallen to about 10\%
for the first two time series, and then becomes too weak to be statistically
significant.  

A comparison of Lorentz factors is also instructive. 
\citet{2011MNRAS.414.2642L} find values of $\sim200-600$, which should be
directly comparable to the asymptotic Lorentz factors in our sample, for
which we obtain similar values, between about 30 and 670 (see
table~\ref{tab:best}), from our observations.  For GRB\,061007, at XRT observation time, the
jet seems to be slowing down, but we still find apparent superluminal
expansion of the blackbody radius, which means that the Lorentz factor is
still significant.  Our proposed model is furthermore supported by
observations such as that of \cite{2012ApJ...757L..31A}, who report a
detection of prompt phase thermal emission for GRB\,110721A, following the
blackbody temperature all the way down to 4.9\,keV.

With this interpretation, our discovery of highly luminous quasi-thermal
components in the soft X-ray emission allows the photospheric prompt
emission model to be explored at much later times and with the more
sensitive narrow-field soft X-ray instruments, and thus potentially with
better statistics and spectral resolution than has so far been the case.

%
%
\section{Conclusions\label{conclusions}}

We have examined a sample of the brightest \emph{Swift} GRBs looking for thermal emission in the XRT data. We find clear evidence for this emission in 8 out of 28 bursts, with an indication that such emission exists in the majority of bursts. We track the temperatures and luminosities of these components over time. We find that several of these thermal components are very luminous (three to four orders of magnitude more luminous than the component in GRB\,060218) and the temperature is high. These facts therefore make SN shock breakout an unlikely explanation in the generality of GRB thermal components, since the components uncovered here are physically similar to other thermal components discovered so far. We find that several of the components have apparent superluminal expansion, indicating that they are clearly expanding relativistically and use late photospheric emission from the jet as a physically well-motivated theory, to allow us to determine Lorentz factors for the bursts. We find the decay rates of the luminosity and temperatures as well as the Lorentz factors to be compatible with values obtained elsewhere from prompt gamma-ray thermal emission. This explanation links the emission observed in the prompt phase to that of later times and is supported by the detection of superluminal motion and the observation that the trends observed in the gamma-ray thermal emission reported extend to the soft X-ray regime and may mark a crucial step in understanding the prompt/early phase of GRB emission.

\begin{acknowledgements}
The Dark Cosmology Centre is funded by the DNRF. MF acknowledges support by the University of Iceland Research fund. We thank P\'all Jakobsson and Gunnlaugur Bj\"ornsson for helpful comments. This work made use of data supplied by the UK Swift Science Data Centre at the University of Leicester.

\end{acknowledgements}

\bibliography{bib_thermal}

\end{document}